\documentclass[amsmath,amssymb,floatfix,prb,showpacs,twocolumn]{revtex4-1}
\usepackage{amsmath,amssymb,natbib,bm,graphicx,url,epsfig}
\usepackage[ansinew]{inputenc}
\usepackage[bookmarks=true]{hyperref}
\usepackage{bm}
\usepackage{color}
\usepackage[justification=raggedright,format=plain,indention=.3cm,font=small]{caption}
\usepackage{subcaption}

\newcommand{\be}{\begin{equation}}
\newcommand{\ee}{\end{equation}}
\newcommand{\bes}{\begin{equation}\begin{split}}
\newcommand{\ees}{\end{split}\end{equation}}
\newcommand{\bea}{\begin{eqnarray}}
\newcommand{\eea}{\end{eqnarray}}

\def\beq{\begin{equation}}
\def\eeq{\end{equation}}
\def\bea{\begin{eqnarray}}
\def\eea{\end{eqnarray}}

\begin{document}

\title{Universal Finite-Size Scaling around Topological Quantum Phase Transitions}
\author{Tobias Gulden$^1$, Michael Janas$^1$, Yuting Wang$^{1}$, and Alex Kamenev$^{1,2}$}

\affiliation{$^1$School of Physics and Astronomy, University of Minnesota, Minneapolis, MN 55455, USA}
\affiliation{$^2$William I. Fine Theoretical Physics Institute, University of Minnesota, Minneapolis, MN 55455, USA}

\date{\today}
\vspace{0.1cm}

\begin{abstract}
The critical point of a topological phase transition is described by a conformal field theory, where finite-size corrections to energy are uniquely related to its central charge. We investigate the finite-size scaling away from criticality and find a scaling function, which discriminates between phases with different topological indexes.  This function appears to be universal for all five Altland-Zirnbauer symmetry classes with non-trivial topology in one spatial dimension. We obtain an analytic form of the scaling function and compare it with numerical results.
\end{abstract}

\maketitle

Since the introduction of topological order in condensed matter physics, the field of topological insulators received constantly growing attention\cite{KaneMele2005,HasanKane,JMoore,QiZhang}. Although non-interacting topological phases were fully classified\cite{AltlandZirnbauer,RyuSchnyder,Stone} and a plethora of topological edge states characterized\cite{HasanKane,QiZhang,Koenig,Zhang,Heiblum,Hasan}, little attention was given so far to finite-size effects around the topological transition. An important question is whether finite-size scaling  is capable to distinguish between topological indexes and may be used as an indicator of the topological nature of the transition. One may also ask whether such scaling is universal or specific to a particular symmetry class, {\em e.g.} sensitive to $\mathbb{Z}$ vs. $\mathbb{Z}_2$ topological index.

In this paper we discuss the finite-size scaling of the ground state energy across topological phase transitions in $1+1$ dimensional models. The critical point in such models is described by a conformal field theory\cite{CFTyellowpages} (CFT). The finite-size, $N$, scaling of the ground state energy $E(N, 0)$ for an {\em open} system at criticality is known\cite{Cardy86,Affleck} to be
\begin{equation}
 E(N,0) = N\,\bar{\epsilon}(0) + b(0) - \frac{c}{N}\, \frac{\pi }{24} +\mathcal{O}(N^{-2}), 
 \label{eq:CFT}
\end{equation}
where $\bar{\epsilon}(0)$ is the average bulk energy per particle, $b(0)$ the size-independent boundary term and argument $(0)$  specifies the exact critical point. Here length is measured in units of lattice spacing and energy in units of the Fermi velocity over the lattice spacing. The $1/N$ term appears to be universal and depends only on  $c$ -- the central charge of the Virasoro algebra\cite{CFTyellowpages}. 

\begin{figure}[h!]
 \includegraphics[width=1\columnwidth]{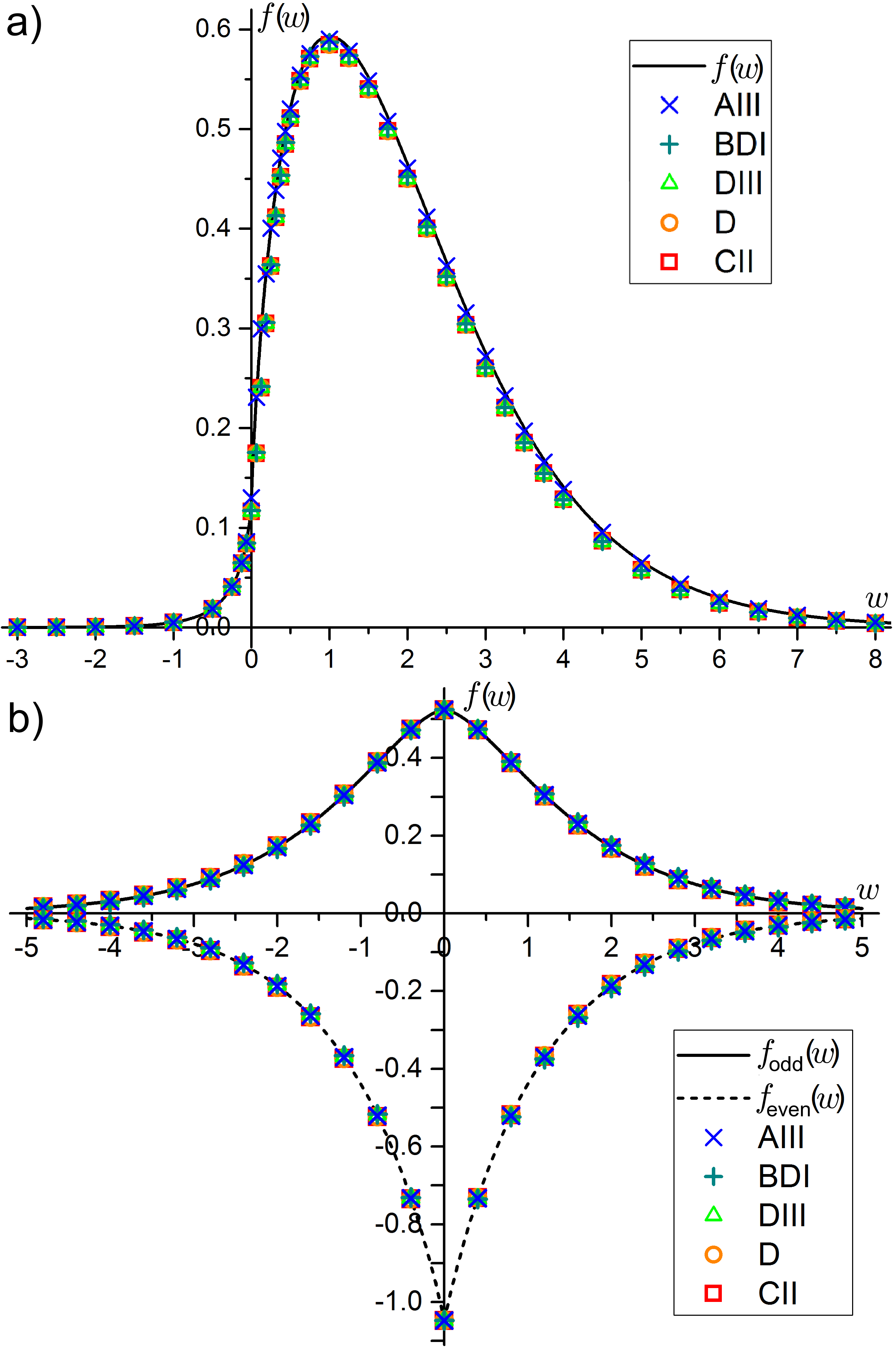}
 \caption{(Color online) Numerical results for $f(w)$, where $w=Nm$ and $N=100$ for 5 topologically non-trivial symmetry classes in one spatial dimension. (a) The case for open boundary conditions which is sensitive to topology, the solid line is the scaling function given by Eq.~\eqref{eq:scaling}. (b) In periodic boundary conditions the results are independent of the topological index and the scaling function Eq.~\eqref{eq:scalingperiodic} is symmetric. There is a difference between even ($f(w)$ negative) and odd ($f(w)$ positive) number of unit cells $N=100,101$.}
 \label{fig:ScalingClasses}
\end{figure}

A relevant perturbation drives the system away from criticality, creating a spectral gap $2m$ and a corresponding  correlation length $\xi=1/m$. Our main observation is that the CFT expansion (\ref{eq:CFT}) may be generalized as
\begin{equation}
 E(N,m) = N\,\bar{\epsilon}(m) + b(m) - \frac{c}{N}\,    f\!\left(Nm \right)+\mathcal{O}(N^{-2})\, ,
 \label{eq:expansion}
\end{equation}
where in the double scaling limit\cite{McCoy}: $N \to\infty$ and $m\to 0$, while $w=Nm=N/\xi = const$, the function $f(w)$, Fig.~\ref{fig:ScalingClasses}, is {\em universal} for all 5 Altland-Zirnbauer symmetry classes with non-trivial topology in 1 spatial dimension (AIII, BDI, DIII, D, CII)\cite{AltlandZirnbauer,RyuSchnyder}. Hereafter we identify $m>0$ with the topological and $m<0$ with non-topological, or lesser topological index, side of the transition. Most notably, the scaling function for open boundary conditions exhibits markedly distinct behavior on the two sides of the topological transition, while for periodic boundary conditions it is symmetric\cite{foot1}. Curiously, a similar scaling function for the entanglement entropy \cite{CardyCalabrese} appears to be symmetric across the topological transition\cite{unpublished}, and it is only the $N$-independent boundary term which is sensitive to the topological index.

One may worry that in the double scaling limit dependence on $m$ of the bulk and boundary terms should not be kept. This is not quite so, because of their singular dependence on the gap. As we explain below
\begin{eqnarray}
\label{eq:bulk}
 \bar{\epsilon}(m) &=& \bar{\epsilon}(0) -\frac{c}{2\pi} \left[m^{2} + \mathcal{O}(m^4)\right] \ln \alpha |m|\,; 
 \\ 
 b(m) &=& b(0)+ \frac{c}{\pi}\left[ m +\mathcal{O}(m^2)\right] \ln \alpha_b |m|, 
 \label{eq:singularEB}
\end{eqnarray}
where $\alpha$ and $\alpha_b$ are non-universal constants. As a result the double scaling limit (\ref{eq:expansion}) for the energy may be equivalently written as  
\begin{equation}
 E(N,w) = N\,\bar{\epsilon}(0) + b(0)+ \frac{c\log N}{2\pi N}\left(2w-w^2 \right) - \frac{c}{N}\,    f_2\!\left(w \right),
 \label{eq:expansion1}
\end{equation}
where ${f_2(w)=f(w)+\frac{1}{2\pi}(2w\log \alpha_b|w| - w^2\log\alpha|w|)}$ incorporates non-universal terms $\sim w$ and $\sim w^2$. Since these latter may be easily subtracted both numerically and analytically, it is preferable to use the expansion (\ref{eq:expansion}) with the fully universal function $f(w)$. 

\begin{figure}[h!]
 \includegraphics[width=1\columnwidth]{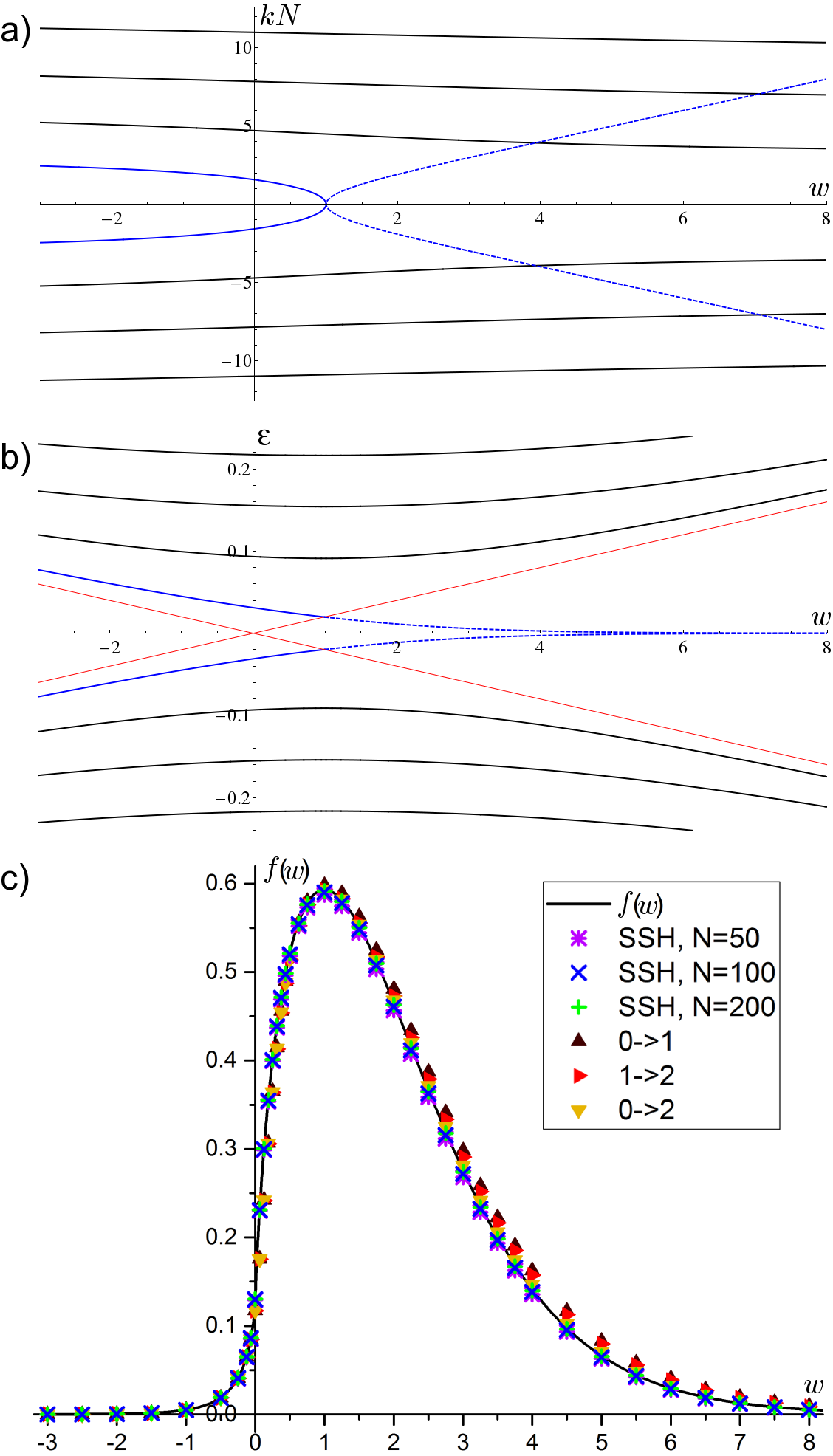}
 \caption{(Color online) SSH model (AIII symmetry class): (a) Visualization of the phase shift across the transition. For a fixed $w=Nm$, a state exists with energy $\epsilon^{\pm}(k)$ if $w=kN\cot(kN)$, Eq.~\eqref{eq:spectrum}. In both limits $w\to\pm\infty$ there are states at $kN=n\pi$, however one pair of states collides at $w=1$ (blue) and obtains imaginary $k$, shown as dashed line. (b) Energy spectrum near the gap for a $N=50$ SSH Hamiltonian (black), with a pair of evolving edge states (blue). The edge state crosses the bulk Dirac cone (thin red) at $w=1$, where the momentum becomes imaginary. (c) Comparison of numerical results for SSH Hamiltonians with the scaling function \eqref{eq:scaling} for system sizes $N=50, 100, 200$ and three transitions in a 2-chain SSH model with topological indexes $Z=0\to1,1\to2,0\to2$.}
 \label{fig:SpectrumSSH}
\end{figure}

{\em Universality of the scaling function:}
Before discussing analytic properties of the scaling function $f(w)$ let us focus on our numerical setup and demonstrate the universal behavior for different symmetry classes. We consider models for all five Altland-Zirnbauer symmetry classes which are topologically non-trivial in one dimension\cite{RyuSchnyder}. To extract the scaling function we use Eq.~\eqref{eq:expansion}. $E(N,m)$ is the sum of all eigenvalues which are obtained by numerical diagonalization. The average energy $\bar{\epsilon}$ is the integral over the entire Brillouin zone of the dispersion relation for all filled bands, which are calculated from the k-space representation of the Hamiltonian. The boundary term may be also calculated analytically (see below) or alternatively approximated by $b(m)\approx E(N,m)-N\bar{\epsilon}(m)$ for some large $N$, say $N=1000$. We have checked that the two ways are in excellent agreement. 

In AIII symmetry class we use the standard Su-Schrieffer-Heeger (SSH) tight-binding Hamiltonian \cite{sshPRL,sshTopol}: 
\begin{equation}
 {\cal H}_{AIII} = \sum_{j=1}^{N} t_1 c_{A,j}^\dagger c_{B,j} + \sum_{j=1}^{N-1} t_2 c_{B,j}^\dagger c_{A,j+1} + h.c.
 \label{eq:AIII}
\end{equation}
Here we choose the gauge for the momentum to have the gap closing at $k=0$, where the dispersion relation reads $\varepsilon(k) = \pm\sqrt{t_1^2 + t_2^2 - 2t_1t_2\cos k}$. For $t_1\neq t_2$ neighboring sites form dimers, where for $t_1>t_2$ all sites are part of a dimer, but for $t_1<t_2$ the two sites at the ends of the chain are unpaired (from now on $m=t_2-t_1$). Thus there are two distinct phases with topological index $Z=0$ or $Z=1$ respectively. In the case of $n$ similar parallel chains the topological index takes values $Z=\{0,1,...,n\}$. Fig.~\ref{fig:SpectrumSSH}c shows results for the scaling function at different system sizes, as well as three different transitions ($Z=0\to1,0\to2,1\to2$) in a system with two parallel chains. The results scale with $w$ as the only parameter, and all transitions agree with the analytic result, Eqs.~\eqref{eq:expansion}, \eqref{eq:scaling}. The other symmetry classes are discussed in the supplemental material\cite{supplement}. Here we only show numerical results in Fig.~\ref{fig:ScalingClasses}, which confirm universality across all five topological classes.

{\em Analytic properties:}
Universality of $f(w)$ function is related to the fact that, similarly to the CFT result (\ref{eq:CFT}), it is fully determined by the vicinity of the critical point. One may thus approximate a near-critical system by the Dirac Hamiltonian, e.g. in AIII symmetry class, ${\cal H}=m\sigma_1 +i\partial_x\sigma_2$, where the Pauli matrices act in the space of $A/B$ sublattices, cf. Eq.~\eqref{eq:AIII}. Assuming that outside of the interval $0<x<N$ the mass is very big and, {\em e.g.}, negative one derives the boundary conditions $\Psi_A(0)=\Psi_B(N)=0$. The quantized values of $k>0$ are given by
\begin{equation}
 \label{eq:spectrum}
 \cos(kN+\delta(k))=0; \quad\quad \tan \delta(k)=\frac{m}{k}=\frac{w}{kN}. 
\end{equation}
As a result the spectrum is determined by the condition $w\equiv Nm = kN\cot (kN)$, plotted in Fig.~\ref{fig:SpectrumSSH}a,  and is given by  $\epsilon^{\pm}(k)=\pm\sqrt{m^2+k^2}$.  At $w=1$ two of its real solutions collide and switch to purely imaginary ones, corresponding to the decaying edge states. Notice that the non-propagating states do not form at $m=0$, as could be naively expected, but rather at $m=1/N$.

Using the argument principle the total groundstate energy is given by 
\begin{equation}
 E(N,m) = \frac{1}{2}\oint \frac{dk}{2\pi i}\epsilon^{-}(k)\, \partial_k \ln\left[\cos(Nk+\delta(k))\right],
 \label{eq:argumentprinciple}
\end{equation}
for the dispersion relation $\epsilon^{-}(k)$ of the filled lower band. The contour runs in the complex $k$-plane encircling all solutions of Eq.~(\ref{eq:spectrum}). The bulk and boundary terms are given by ${N\bar\epsilon+b = \int (dk/2\pi)\epsilon^{-}(k)[N+\partial_k\delta(k)]}$, where $N+\partial_k\delta(k)$ are bulk and boundary parts of the continuous density of states. To find the scaling function $f(w)$ one subtracts $N\bar\epsilon +b$ from Eq.~(\ref{eq:argumentprinciple}), deforms the integration contour to run along the branch cut of $\sqrt{m^2+k^2}$ and rescales the integration variable as $z=ikN$. As a result, one finds
\begin{equation}
 f(w) = -\int_{|w|}^{\infty} \frac{dz}{\pi}\sqrt{z^2-w^2}\, \partial_z \ln\left[1+e^{-2z-2\delta_w(z)}\right],
 \label{eq:scaling}
\end{equation}
where $\delta_w(z)=-\tanh^{-1}(w/z)$. (For more detail on the derivation see the supplemental material\cite{supplement}.) This expression is plotted in Figs.~\ref{fig:ScalingClasses}a, \ref{fig:SpectrumSSH}c as a full line and is in good agreement with the numerical data.

Before discussing analytic properties of this scaling function let us add a couple of remarks: (i) though the derivation  was given for the model in symmetry class AIII,  the same logic works for the other symmetry classes. One needs to subtract proper model-dependent bulk and boundary parts, but the scaling term is only determined by the vicinity of the Dirac point and remains unchanged; (ii) a similar derivation may be applied to the case of periodic boundary conditions. In the gauge chosen after equation \eqref{eq:AIII} periodic boundary conditions give $\Psi(N)=(-1)^N\Psi(0)$. In this case the quantization condition \eqref{eq:spectrum} changes to $\cos(kN)=(-1)^N$. After subtracting the bulk energy (there is no boundary term in this case) and following the same steps one arrives at a similar scaling function:
\begin{eqnarray}
 f(w) &=& -2\int_{|w|}^{\infty} \frac{dz}{\pi}\sqrt{z^2-w^2}\, \partial_z \ln\left[1-(-1)^Ne^{-z}\right]\nonumber\\
      &=& -\frac{2w}{\pi}\sum_{j=1}^\infty\frac{K_1(jw)}{j}(-1)^{jN},
 \label{eq:scalingperiodic}
\end{eqnarray}
where $K_1(x)$ is the modified Bessel function. This function $f(w)$ is manifestly symmetric across the topological phase transition, as it must be  for periodic boundary conditions. However even in the scaling limit $N\to\infty$ it is dependent on parity of $N$, see also Fig.~\ref{fig:ScalingClasses}b. The difference may be attributed to the level crossing at the gap closing point $k=0$ and $w=0$ for even $N$, explaining $\sim |w|$  non-analytic behavior of the scaling function. For odd $N$, all levels undergo avoided crossings and the scaling function is free from such non-analyticity. The CFT result\cite{Cardy86,Affleck} predicts $f(0)=\pi/6$, which agrees with the case for odd $N$, while for even $N$ we obtain $f(0)=-\pi/3$. 

Returning to a system with open boundary conditions, at small $|w|\ll 1$ Eq.~(\ref{eq:scaling}) leads to:
\begin{equation}
 f(w)\approx \frac{\pi}{24} + \frac{1}{2\pi}\left(-2w + w^2 \right) \ln|w| + \dots . 
 \label{eq:small-Lm}
\end{equation}
The first term here is in agreement with the CFT limit (\ref{eq:CFT}). The subsequent terms ensure that $f_2(w)$ function, defined after Eq.~(\ref{eq:expansion1}), is analytic. Indeed, at any finite $N$ the ground state energy $E(N,m)$ and all its derivatives must be non-singular at $m=0$. To derive the second term in Eq.~(\ref{eq:small-Lm}) one may employ  monodromy transformation, which rotates complex $w$ in a small circle around zero\cite{Zoladek}. Upon such transformation the right hand side of Eq.~(\ref{eq:scaling}) picks up a contribution given by a closed contour integral around a branch cut $-|w|<z<|w|$ times the number of revolutions. Calculation of such an integral leads to  $i(-2w+w^2)$, implying that $f(w)$ must have logarithmic branch cut terminating at $w=0$ with the discontinuity across it given by this value. Hence Eq.~(\ref{eq:small-Lm}). We note in passing that in addition to such logarithmic branch cut, $f(w)$ function has an infinite sequence of square root branch cuts along the imaginary axis of $w$.

At large argument $|w|\gg 1$, i.e. $N\gg|\xi|$, the finite-size corrections decay exponentially. Remarkably the rate of the decay appears to be sensitive to the topology: 
\begin{equation}
 f(w)\approx \left\{  \begin{array}{ll}
 \frac{1}{16\sqrt{\pi}}|w|^{-\frac{1}{2}}e^{-2|w|} & \mbox{trivial}\,\quad \quad\, w\ll -1; \\
 2w e^{-w} \quad& \mbox{topological}\,\, w\gg 1.
 \end{array}
 \right.
 \label{eq:large-Lm}
\end{equation}
(In $\mathbb{Z}$ symmetry classes, the two lines may be attributed to $Z$ and $Z+1$ topological index.) The fact that on the topological side the scaling function decays half as fast as on the trivial side may be associated with the appearance of the edge states in the middle of the gap and effectively cutting the gap in half. In fact, the purely imaginary solution of $w=kN\cot(kN)$ at  $w\gg 1$ gives the energy of the edge states as $\epsilon=\pm 2w e^{-w}/N$. This is identical to the asymptotic of $f(w)/N$ on the topological side of the transition, Eq.~\eqref{eq:large-Lm}, indicating that the latter originates solely from the edge state. In the case of periodic boundary conditions the large $w$ asymptotic is $f(w) = -(-1)^N\sqrt{2/\pi}|w|^{1/2}e^{-|w|}$, which is different from both sides of the transition, Eq.~(\ref{eq:large-Lm}), in the open boundary condition case.  

Furthermore note that there develops a peak at $w=N/\xi=1$ on the topological side (cf. Fig.~\ref{fig:ScalingClasses}). At this point there is a crossover between the regime of the correlation length being larger than the system size to smaller than the system size. In other words, here the two edge states at opposite ends transform from being delocalized and correlated to localized modes, i.e. the topological transition happens when $m=1/N$. This manifests itself in Fig.~\ref{fig:SpectrumSSH} as the point where two momenta become imaginary.

{\em Conclusions and outlook:}
In conformal field theories the $N^{-1}$ term in energy is universal and only depends on the central charge of the Virasoro algebra\cite{Cardy86,Affleck}. Here we find that in the case of topological phase transitions this term naturally extends into a scaling function, depending only on the ratio of the system size to the correlation length. Furthermore this scaling function is universal for all topologically non-trivial classes of non-interacting fermions in one spatial dimension. While the scaling function for energy appears to be sensitive to the topological nature of the transition, this is by no means the common situation. For example, the finite-size scaling function of the entanglement entropy away from the critical point\cite{CardyCalabrese}, appears to be symmetric across the topological transition\cite{unpublished} (there is still an asymmetric size-independent boundary term).      

It is natural to ask whether the scaling behavior changes with interactions, especially for models with central charge different from $c=1$ and $c=1/2$ considered here. Another direction to explore is relation of the scaling function to the theory of integrable systems\cite{integrable}. In particular if it may be expressed in terms of solutions of Painleve equations, as it happens in e.g. the Ising model \cite{McCoy}.  As mentioned above, the non-analytic contributions to the finite-size scaling near $w=0$ are related to the monodromy, i.e. discontinuity across the branch cut terminating at $w=0$, which happens to be simply a second order polynomial in $w$.  An open question is if the full $f(w)$-function may be recovered from the monodromy data, specified for all of its branch cuts, through the solution of a Riemann-Hilbert problem\cite{Zoladek}. 

We are grateful to A. Abanov and I. Gruzberg for valuable discussions. The work was supported by NSF grant DMR1306734.


\section{Supplemental Material to Universal Finite-Size Scaling around Topological Quantum Phase Transitions}
\paragraph{Different symmetry classes:} In this section we introduce model systems for all topologically non-trivial symmetry classes in one dimension\cite{RyuSchnyder}. As mentioned in the main text, the Su-Schrieffer-Heeger model\cite{sshPRL,sshTopol} belongs in {\bf AIII} symmetry class, it is a bipartite fermionic model with tight-binding Hamiltonian
\begin{equation}
 \mathcal{H}_{AIII} = \sum_{j=1}^{N} t_1 c_{A,j}^\dagger c_{B,j} + \sum_{j=1}^{N-1} t_2 c_{B,j}^\dagger c_{A,j+1} + h.c.
\end{equation}
and dispersion relation $\varepsilon(k) = \pm\sqrt{t_1^2 + t_2^2 - 2t_1t_2\cos k}$. For $t_1\neq t_2$ neighboring sites form dimers, where for $t_1>t_2$ all sites are part of a dimer, but for $t_1<t_2$ the two sites at the ends of the chain are unpaired. These are the two distinct phases with topological index $Z=0$ or $Z=1$ respectively. The central charge is $c=1$.

The spinless Kitaev chain\cite{Kitaev,Alicea} falls into symmetry class {\bf BDI}. The Hamiltonian is
\begin{equation}
 \mathcal{H}_{BDI} = -2\mu\sum_{j=1}^N c_j^\dagger c_j -\sum_{j=1}^{N-1} \left(tc_j^\dagger c_{j+1} + \Delta c_jc_{j+1}\right) + h.c.,
 \label{eq:BDI}
\end{equation}
where $\mu$ is the chemical potential, $t$ a hopping parameter and $\Delta$ the pair creation amplitude. For $|\mu|<t$ the system is topological and Majorana edge modes appear, the gap size gives $m=t-|\mu|$. The central charge is $c=1/2$.

A generalization to spin-$\frac{1}{2}$ fermions with p-wave pairing falls into symmetry class {\bf DIII}\cite{SchnyderNancy}:
\begin{eqnarray}
 \label{eq:DIII}
 \mathcal{H}_{DIII} &=& -2\mu\sum_{\sigma,j=1}^N c_{j,\sigma}^\dagger c_{j,\sigma} 
                        +t\sum_{\sigma,j=1}^{N-1} c_{j,\sigma}^\dagger c_{j+1,\sigma}\\\nonumber
                    & & +\sum_{j=1}^{N-1} \left(\Delta_1c_{j,\uparrow}^\dagger c_{j+1,\uparrow}^\dagger 
                                                       +\bar{\Delta}_1c_{j,\downarrow}^\dagger c_{j+1,\downarrow}^\dagger\right)\\\nonumber
                    & & +\Delta_2 \sum_{\sigma\neq\sigma',j=1}^{N-1}-ic_{j,\sigma}^\dagger c_{j+1,\sigma'}^\dagger +h.c.
 \\\nonumber
\end{eqnarray}
Here $\mu$ is the on-site potential, $t$ hopping parameter and $\Delta_1,\Delta_2$ p-wave pairing fields. For $|\mu|<t$ the system is in a topological state, where the gap size gives $m=t-|\mu|$. Due to Kramer's degeneracy all Majorana modes appear in pairs, thus $c=1$.

Introducing an additional Zeeman splitting of the on-site potential breaks time reversal symmetry and gives a system in {\bf D} symmetry class. Compared to the DIII Hamiltonian we replace the on-site term as follows:
\begin{equation}
 \mathcal{H}_D: -2\mu\sum_{\sigma,j=1}^N c_{j,\sigma}^\dagger c_{j,\sigma} \to -2\sum_{j=1}^N \mu_1c_{j,\uparrow}^\dagger c_{j,\uparrow}
                                                                                             +\mu_2c_{j,\downarrow}^\dagger c_{j,\downarrow}.
 \label{eq:D}
\end{equation}
The system is topological if $t\in(\mu_1,\mu_2)$, so there are two topological critical points with $c=1/2$. Focussing on the latter the size of the gap yields $m=t-\mu_2$.

The {\bf CII} symmetry class may be represented by a bipartite tight-binding model for spin-$\frac{1}{2}$ fermions. 
\begin{eqnarray}
 \mathcal{H}_{CII} &=& \mu \sum_{\sigma,j=1}^N (a_{j,\sigma}^\dagger a_{j,\sigma} - b_{j,\sigma}^\dagger b_{j,\sigma})\\\nonumber
                   & & +\frac{t_1}{2} \sum_{\sigma,j=1}^{N-1} (a_{j,\sigma}^\dagger a_{j+1,\sigma} - b_{j,\sigma}^\dagger b_{j+1,\sigma})\\\nonumber
                   & & +\frac{t_2}{2i} \sum_{j=1}^{N-1} (- a_{j,\uparrow}^\dagger b_{j+1,\downarrow} + a_{j+1,\uparrow}^\dagger b_{j,\downarrow}\\\nonumber
 & & \textcolor{white}{+\frac{t_2}{2i} \sum_{j=1}^{N-1}} + a_{j,\downarrow}^\dagger b_{j+1,\uparrow} - a_{j+1,\downarrow}^\dagger b_{j,\uparrow}) + h.c.,
 \label{eq:CII}
\end{eqnarray}
where $\mu$ is the on-site potential and $t_1,t_2$ are hopping parameters. The critical point is at $|\mu|=t_1$, with $m=t_1-\mu,$ and the transition is described by a conformal field theory with $c=1$.

\begin{figure}[h!]
 \includegraphics[width=1\columnwidth]{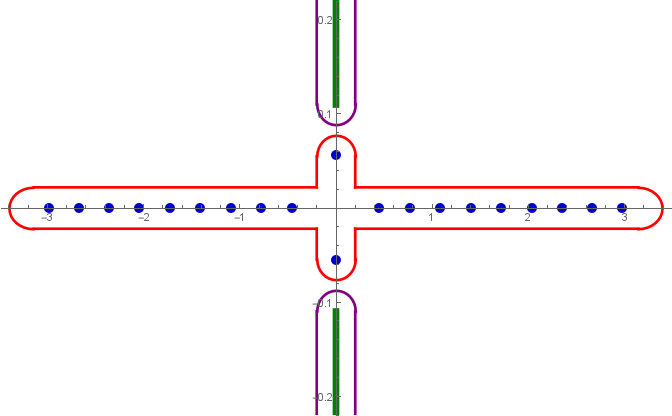}
 \caption{Integration contour in the complex $k$ plane for $N=10$ and $w=1.1$. The contour (red) encloses all zeroes of the argument of the logarithm (blue) and is to be taken counter-clockwise. After a change of variable $z=Nk$ and taking the scaling limit $N\to\infty$ the contour runs along the entire real line, above and below. These two parts of the contour  are now deformed (purple) to run along the branch cuts of $\sqrt{m^2+k^2}$ (green).}
 \label{fig:contour}
\end{figure}

\paragraph{Derivation of the scaling function} In this section we want to provide more detail on the derivation of the scaling function \eqref{eq:scaling}. The total energy of the filled lower band is the sum of all states, satisfying  $\cos(kN+\delta(k))=0;\quad \tan \delta(k)=\frac{m}{k}=\frac{w}{kN}$, Eq.~(7) in the main text, substituted in the lower band dispersion relation $\epsilon^{-}(k) = -\sqrt{t_1^2 + t_2^2 - 2t_1t_2\cos k}$. This sum may be converted into an integral with the argument principle:
\begin{equation}
 E(N,m) = \frac{1}{2} \oint \frac{dk}{2\pi i}\epsilon^{-}(k)\, \partial_k \ln\left[\cos(Nk+\delta(k))\right],
 \label{eq:argumentprinciple1}
\end{equation}
where the contour runs in the complex $k$-plane encircling all solutions of the quantization condition, see Fig.~\ref{fig:contour}. The prefactor $1/2$ is introduced to avoid double counting. The two pieces of the integration contour above and below the real axis yield the same contribution, in the following we treat these two pieces independently. The contributions proportional to $N^1$ and $N^0$ are the bulk and boundary terms, respectively. Writing the cosine as sum of two exponentials one can factor out these leading contributions to get
\begin{eqnarray}
 N\bar\epsilon + b &=& \int (dk/2\pi)\epsilon^{-}(k)[N+\partial_k\delta(k)]\\\nonumber
                   &=& \pm\int (dk/2\pi i)\epsilon^{-}(k)\partial_k \ln e^{\pm i(Nk+\delta(k))},
\end{eqnarray}
where the $\pm$ sign is chosen differently above and below the real line so that the remaining exponential term vanishes as $k\to\pm i\infty$. Now one subtracts $N\bar\epsilon+b$ from Eq.~\eqref{eq:argumentprinciple1} to find the scaling function $f(w)$,
\begin{equation}
 -\frac{f(w)}{N} = \pm\int\frac{dk}{2\pi}\epsilon^{-}(k)\, \partial_k \ln\left[1+e^{\mp2i(Nk+\delta(k))}\right].
\end{equation}
In the scaling limit $N\to\infty$ integration runs along the entire real line and can be seen as closed at infinity. The approximation to the dispersion relation near the Dirac point, $\epsilon^{-}(k) = \sqrt{m^2+k^2}$, has a branch cut along the imaginary axis starting at $k=\pm im$. Now one deforms the integration contour to run along this branch cut and redefines the integration variable as $z=\pm ikN$ and introduces $w=Nm$. As a result, in the scaling limit $N\to\infty$ one finds
\begin{equation}
 f(w) = -\int_{|w|}^{\infty} \frac{dz}{\pi}\sqrt{z^2-w^2}\, \partial_z \ln\left[1+e^{-2z-2\delta_w(z)}\right],
\end{equation}
where $\delta_w(z)=-\tanh^{-1}(w/z)$, which is Eq.~(9) of the main text.


\begin{thebibliography}{99}
\bibitem{foot1} Notice that the average energy $\bar\epsilon(m)$ is a symmetric function across the transition, while the boundary term $b(m)$ is sensitive to the presence of the edge states and is therefore asymmetric. In periodic boundary conditions the two sides of the transition are physically equivalent.
\end{thebibliography}
\end{document}